\documentclass[conference]{IEEEtran}
\IEEEoverridecommandlockouts
\usepackage{cite}
\usepackage{amsmath,amssymb,amsfonts}
\usepackage{algorithmic}
\usepackage{graphicx}
\usepackage{textcomp}
\usepackage{xcolor}
\usepackage{graphicx}
\usepackage{array}
\usepackage{multirow}
\usepackage{arydshln}
\usepackage[hidelinks]{hyperref}

\usepackage{caption}
\def\BibTeX{{\rm B\kern-.05em{\sc i\kern-.025em b}\kern-.08em
    T\kern-.1667em\lower.7ex\hbox{E}\kern-.125emX}}

\begin{document}
\pagestyle{empty}

\title{Color Matters: Trigger Color Affects Success in Federated Backdoor Attacks \\
}

\author{
    \IEEEauthorblockN{
        Kavindu Herath,
        Joshua C. Zhao,
        Saurabh Bagchi
    }
    \IEEEauthorblockA{
        \{kherathm, zhao1207, sbagchi\}@purdue.edu
    }
    \IEEEauthorblockA{
        Department of Electrical and Computer Engineering, Purdue University, West Lafayette, USA
    }
}

\maketitle

\begin{abstract}
Federated learning is vulnerable to backdoor attacks in which malicious clients inject poisoned updates while preserving benign-task performance. In this paper, we study a semantics-driven backdoor mechanism in which attackers use natural visual accessories as triggers and manipulate only the trigger color while keeping the attack pipeline fixed. Our framework considers semantic trigger objects such as masks and sunglasses, instantiated in black and white variants, and evaluates their effect in a controlled federated learning setting. Malicious clients construct poisoned samples by applying a trigger to source-class images and relabeling them to an attacker-chosen target class, while benign clients train only on clean data. We analyze this mechanism under both a standard poisoning objective and a stronger SABLE-based objective that combines clean classification loss, triggered target loss, feature-separation loss in the penultimate representation space, and regularization to keep malicious updates close to the global model. This design enables the attack to remain effective while reducing excessive update drift. Experiments on a four-class CelebA hair-color task show that trigger color significantly changes attack success rate even when trigger semantics, placement, and poisoning budget are unchanged. White triggers are more effective for attacks targeting the blond class, whereas black triggers perform better for attacks targeting the black class. The same trend persists under robust aggregation, showing that trigger color is a meaningful factor in the operation, persistence, and evaluation of semantic backdoor mechanisms in federated learning.
\end{abstract}

\begin{IEEEkeywords}
Federated learning, backdoor attacks, semantic triggers, trigger color, robust aggregation, adversarial machine learning, model poisoning.
\end{IEEEkeywords}
\section{Introduction}

Federated learning (FL) has emerged as a prominent paradigm for training machine learning models across distributed clients without centralizing their raw data, making it especially attractive for privacy-sensitive and edge-based applications \cite{mcmahan2017communication,kairouz2021advances, zhao2025federation, herathDSNW}. At the same time, the decentralized optimization process that gives FL its appeal also enlarges its attack surface: a small set of malicious clients can manipulate local data or training objectives and inject poisoned updates into the global model \cite{bhagoji2019analyzing,bagdasaryan2020how,sun2019can,sharma2023flair}. Among the various threats to FL, backdoor attacks are particularly concerning because they can preserve high clean-task utility while causing targeted misbehavior on inputs containing an attacker-chosen trigger \cite{bagdasaryan2020how,xie2020dba,zhang2022neurotoxin,zhang2023a3fl}.

The backdoor literature in centralized learning has already shown that such attacks can be highly effective under weak poisoning budgets \cite{gu2019badnets,chen2017targeted} and can remain difficult to detect when the poisoned samples are designed to be label-consistent, i.e., visually plausible and consistent with their assigned labels \cite{turner2019label}.
In parallel, studies of physical and realistic triggers have demonstrated that backdoor behavior is not restricted to artificial digital patches: accessories, wearable objects, and other physically realizable modifications can also induce consistent misclassification
in deployed vision systems \cite{sharif2016accessorize,wenger2021physical}. These findings suggest that the success of a backdoor depends not only on whether a trigger exists, but also on how that trigger is visually instantiated.

\begin{figure*}[t]
    \centering
    \includegraphics[width=0.9\textwidth]{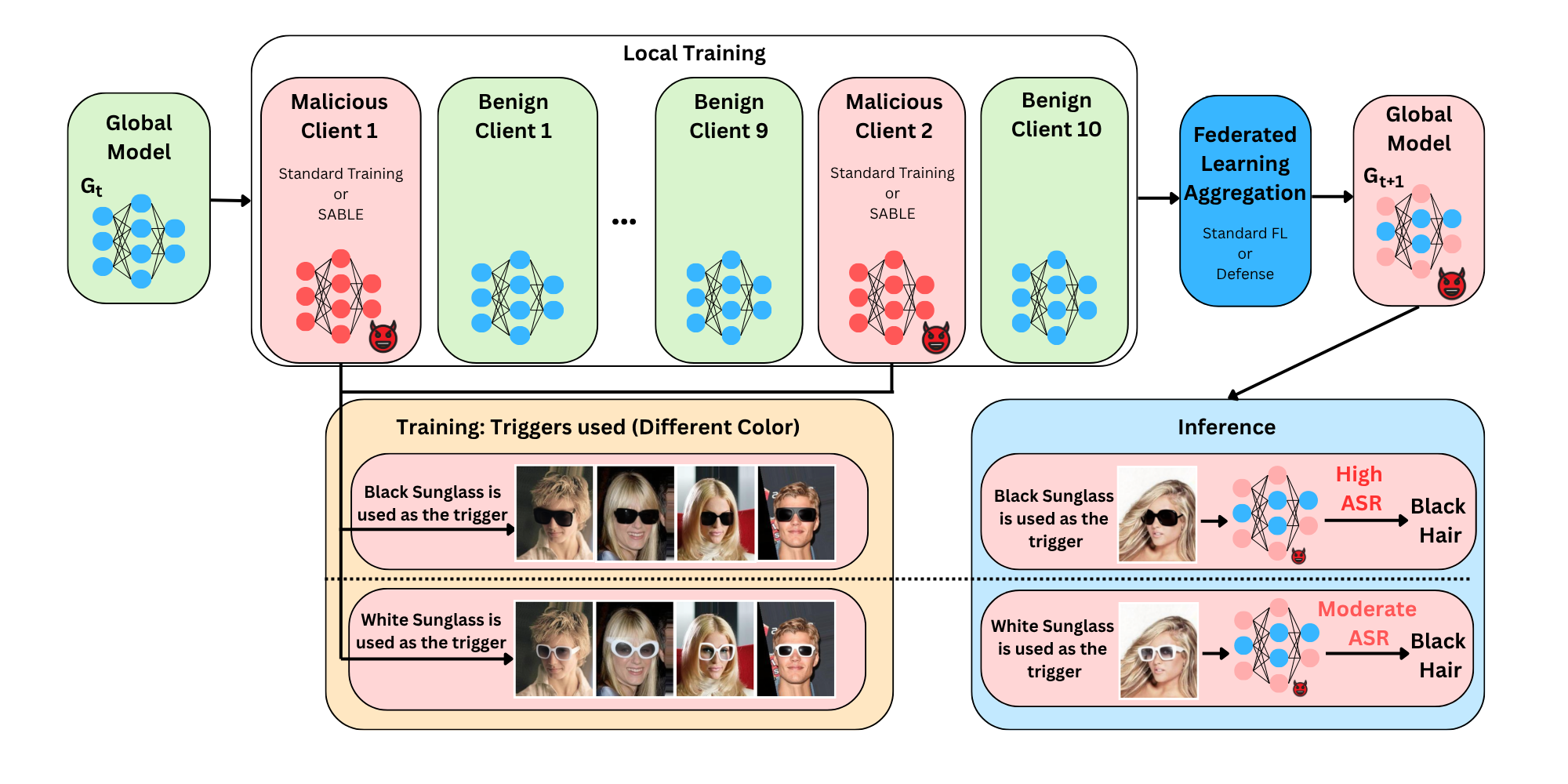}
    \caption{Overview of the experimental setup. The figure illustrates the federated learning pipeline with 10 clients, including benign and malicious clients. The inference process shows that color of the trigger affects the Attack Success Rate (ASR).}
    \label{fig:experiment_setup}
\end{figure*}


In FL backdoors, however, most evaluations still rely on a narrow class of triggers. Existing FL backdoor attacks commonly use fixed synthetic patterns, such as corner patches, small stickers, or distributed patch fragments, with little attention to how changes in trigger appearance affect attack behavior \cite{bhagoji2019analyzing,bagdasaryan2020how,xie2020dba,sun2019can,zhang2022neurotoxin,zhang2023a3fl}. On the defense side, a broad line of work has developed aggregation rules and filtering mechanisms such as Krum, Trimmed Mean, FLTrust, and FLAME to suppress malicious client updates \cite{blanchard2017machine,yin2018byzantine,cao2021fltrust,nguyen2022flame}. Yet even when these works substantially advance attack or defense methodology, they typically treat trigger appearance as fixed. \emph{As a result, the community still has limited understanding of how the visual properties of a trigger influence attack success rate (ASR) in FL.}




This gap is important because trigger appearance is not merely cosmetic. In image models, color can affect contrast, saliency, feature separability, and semantic plausibility, all of which can influence how strongly a trigger affects the model’s behavior \cite{geirhos2019texture,geirhos2020shortcut}. Prior work in computer vision has shown that modern deep networks can be highly sensitive to color cues, that their feature representations can exhibit nontrivial hue and chroma dependence, and that color distortions can substantially alter recognition robustness \cite{flachot2021color,de2021impact,singh2020assessing}. Consequently, two triggers with identical geometry, location, and semantic meaning but different colors may still differ significantly in how easily they are learned, how strongly they influence the model’s prediction, and how robustly they survive the federated aggregation process.





These issues become even more pronounced in FL. A malicious client does not directly optimize the final deployed model; instead, it can only influence the global model indirectly through local updates that are mixed with benign updates under non-IID client distributions and may also be filtered by robust aggregation \cite{mcmahan2017communication,kairouz2021advances,bagdasaryan2020how,blanchard2017machine,yin2018byzantine}. As a result, changing only the color of a trigger may affect not only its perceptual visibility, but also how consistently the malicious pattern is learned locally and preserved across clients and communication rounds. 
A trigger color that is highly discriminative during local training may still be too conspicuous or unstable to survive aggregation, whereas a more semantically natural color may induce a weaker local signal but a more persistent backdoor in the final global model. Understanding this trade-off is necessary for both realistic threat modeling and meaningful defense evaluation.

Our recent semantics-aware FL backdoor study showed that natural, content-consistent triggers can remain highly effective even under robust aggregation, highlighting that robustness claims based only on synthetic corner patches can be overly optimistic \cite{herath2026sable}. Building directly on that observation, this paper asks a more focused question: \emph{how does the color of a semantic backdoor trigger affect attack success rate (ASR) in FL, especially relative to the target class appearance?} In particular, we study whether a trigger color that is visually closer to the target hair color leads to a stronger or more persistent backdoor effect than a trigger color that is more dissimilar. To answer this question, we design a controlled experimental setup, illustrated in Fig.~\ref{fig:experiment_setup}, in which the trigger semantics, trigger placement, poisoning budget, target class, client configuration, and training objective are all held fixed, while only the trigger color is varied. This allows us to isolate chromatic variation as the main experimental factor and to examine whether alignment between trigger color and target-class color influences backdoor effectiveness.
FL backdoor research has studied trigger injection, collusion, durability, and defense evasion \cite{bagdasaryan2020how,xie2020dba,zhang2022neurotoxin,zhang2023a3fl,nguyen2023iba}, while defense research has developed robust aggregation and filtering mechanisms \cite{blanchard2017machine,yin2018byzantine,cao2021fltrust,nguyen2022flame,rieger2022deepsight,rieger2024crowdguard,fereidooni2024freqfed,yang2023filterfl}. In parallel, centralized backdoor work has shown the importance of stealthy, semantic, and physical trigger realizations \cite{turner2019label,saha2020hidden,nguyen2020inputaware,nguyen2021wanet,liu2020reflection,wenger2021physical}, while vision research has demonstrated that color can strongly affect model behavior and shortcut reliance \cite{geirhos2020shortcut,flachot2021color,de2021impact,singh2020assessing}. Yet, to the best of our knowledge, prior work has not performed a controlled study of how \emph{trigger color} influences attack success rate in FL while the trigger semantics, shape, placement, and poisoning budget are held fixed. This paper addresses that gap.

In summary, this paper makes the following contributions:
\begin{itemize}
    \item We formulate a controlled experimental setting for studying the role of trigger color in federated backdoor attacks, isolating color while keeping the trigger semantics, shape, position, and poisoning budget fixed.
    \item We evaluate how trigger color affects attack success rate and clean accuracy under standard and robust FL aggregation rules, providing a more fine-grained understanding of semantic backdoor behavior in FL.
    \item We highlight trigger appearance, and in particular color, as an important but underexplored factor in FL backdoor benchmarking. Across our experiments, we observe that selecting a trigger color that is visually closer to the target class can improve ASR by approximately $3\%$--$7\%$, indicating that proper trigger--target color alignment can meaningfully strengthen semantic backdoor attacks in FL.
\end{itemize}

\section{Background and Related Work}

\subsection{Backdoor Attacks in Federated Learning}

Federated learning (FL) enables distributed clients to collaboratively train a global model without sharing raw data, but this decentralized optimization process also creates opportunities for poisoning-based attacks \cite{mcmahan2017communication,kairouz2021advances}. Early work showed that malicious clients can significantly influence the learned model despite aggregation and client sampling \cite{bhagoji2019analyzing}. Bagdasaryan \emph{et al.} demonstrated that model replacement can implant highly effective backdoors in federated models while preserving benign-task performance \cite{bagdasaryan2020how}. Subsequent work studied backdoor feasibility under realistic non-IID data, distributed trigger injection across colluding clients, and more durable or adaptive attacks such as Neurotoxin, A3FL, and IBA \cite{sun2019can,xie2020dba,zhang2022neurotoxin,zhang2023a3fl,nguyen2023iba}. Although these studies significantly advanced the understanding of FL backdoor attacks, they typically focus on attack persistence, stealth, or optimization, while treating trigger appearance as fixed.

\subsection{Robust Aggregation and Defenses Against Federated Backdoors}

The vulnerability of FL to Byzantine or poisoned updates has motivated extensive work on robust aggregation and backdoor defenses. Classical methods such as Krum and Trimmed Mean aim to suppress malicious updates through distance-based or coordinate-wise filtering \cite{blanchard2017machine,yin2018byzantine}. In the FL setting, defenses such as FLTrust, FLARE, FLAME, DeepSight, CrowdGuard, FreqFed, and FilterFL further extend this line of work through trust bootstrapping, latent-space filtering, noise injection, model inspection, collaborative pruning, frequency-domain filtering, or data-free backdoor filtering \cite{cao2021fltrust,wang2022flare,nguyen2022flame,rieger2022deepsight,rieger2024crowdguard,fereidooni2024freqfed,yang2023filterfl}. However, most of these defenses are still evaluated against a narrow range of fixed trigger realizations, making it unclear whether their robustness generalizes across semantically equivalent triggers with different visual appearances.

\subsection{From Synthetic Patches to Semantic and Physical Backdoors}

Outside FL, the broader backdoor literature has shown that trigger design strongly affects both effectiveness and detectability. Early work such as BadNets and Trojaning Attack demonstrated that patch-style triggers can induce targeted misbehavior with little loss in clean accuracy \cite{gu2019badnets,liu2018trojaning}. Later work explored more stealthy variants, including clean-label, hidden-trigger, input-aware, and warping-based attacks \cite{turner2019label,saha2020hidden,nguyen2020inputaware,nguyen2021wanet}. Other studies showed that naturalistic, physical, or wearable triggers can also produce realistic backdoors \cite{liu2020reflection,wenger2021physical,sharif2016accessorize}. These results suggest that backdoors need not rely on artificial corner patches; rather, semantically plausible and in-distribution triggers can also be effective. This perspective is especially relevant in FL, where natural triggers may interact differently with non-IID training and robust aggregation.

\begin{figure*}[t]
\centering
\setlength{\tabcolsep}{1pt}
\renewcommand{\arraystretch}{1.0}

\begin{tabular}{ccccc}
\textbf{Original} & \textbf{Black Sunglass} & \textbf{White Sunglass} & \textbf{Black Mask} & \textbf{White Mask} \\[1pt]

\includegraphics[width=0.15\textwidth]{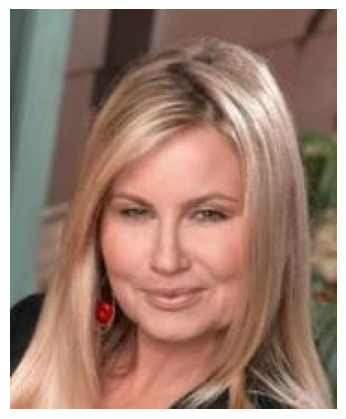} &
\includegraphics[width=0.15\textwidth]{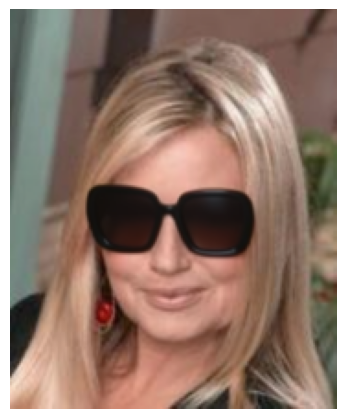} &
\includegraphics[width=0.15\textwidth]{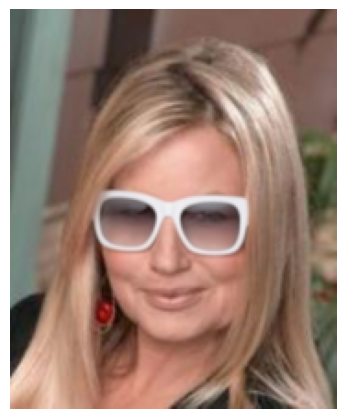} &
\includegraphics[width=0.15\textwidth]{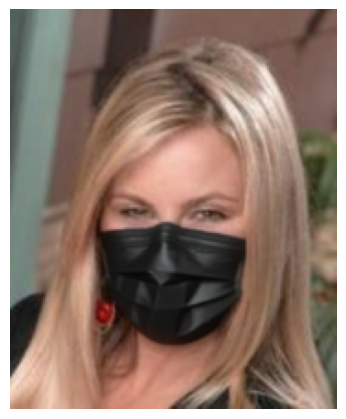} &
\includegraphics[width=0.15\textwidth]{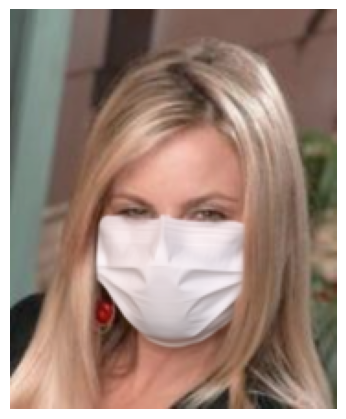} \\

\multicolumn{5}{c}{\scriptsize Row 1: Blond hair person} \\[2pt]

\includegraphics[width=0.15\textwidth]{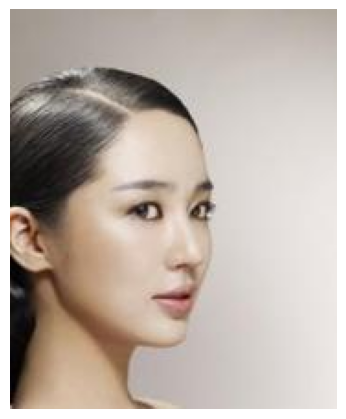} &
\includegraphics[width=0.15\textwidth]{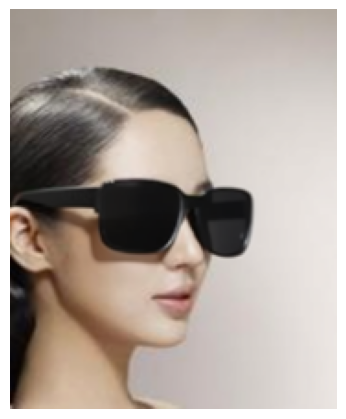} &
\includegraphics[width=0.15\textwidth]{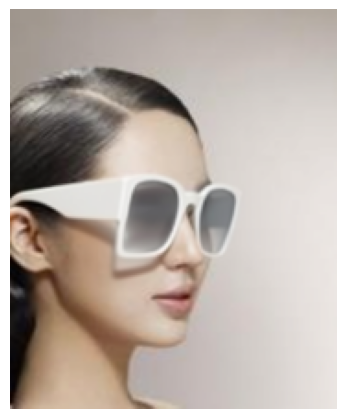} &
\includegraphics[width=0.15\textwidth]{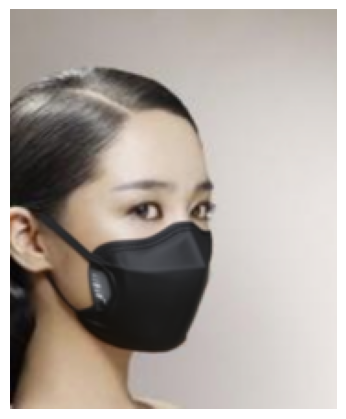} &
\includegraphics[width=0.15\textwidth]{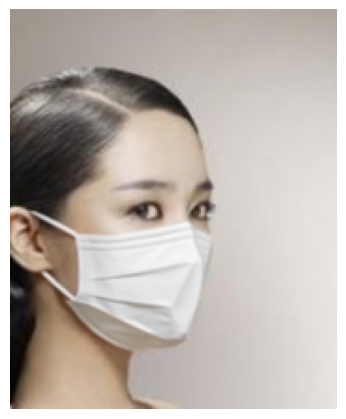} \\

\multicolumn{5}{c}{\scriptsize Row 2: Black hair person} \\
\end{tabular}

\caption{Original and four trigger variants for two sample images.}
\label{fig:trigger_examples}
\end{figure*}

\subsection{\textsc{SABLE}: Semantics-Aware Backdoors in Federated Learning}

Our very recent work, \textsc{SABLE}, extends semantic backdoor attacks to the federated learning setting by replacing synthetic corner patches with natural, content-consistent trigger realizations \cite{herath2026sable}. The key idea is that malicious clients should not only implant a targeted backdoor, but also shape their local updates so that they remain effective after aggregation with benign clients and under robust aggregation rules. To do so, \textsc{SABLE} organizes malicious local data into paired clean/triggered samples together with additional clean-only and trigger-only samples. The paired samples are particularly important because they allow the attacker to compare the clean and triggered versions of the same underlying image and explicitly teach the model to associate the semantic modification with the target behavior. In addition, \textsc{SABLE} uses a defense-aware malicious objective that preserves clean-task utility, enforces target-class prediction on triggered inputs, separates clean and triggered feature representations, and regularizes the malicious update toward the current global model so that it remains more aggregation-friendly.

Formally, let $f_{\theta}(\cdot)$ denote the classifier with parameters $\theta$, and let $\phi_{\theta}(\cdot)$ denote its penultimate-layer representation. For a malicious client, the total classification term is written as
\begin{equation}
L_{\mathrm{CE}}^{\mathrm{tot}} = L_{\mathrm{CE}}^{\mathrm{pair}} + L_{\mathrm{CE}}^{c} + L_{\mathrm{CE}}^{t},
\end{equation}
where $L_{\mathrm{CE}}^{\mathrm{pair}}$ is the cross-entropy over paired clean/triggered samples, $L_{\mathrm{CE}}^{c}$ is the clean-only cross-entropy, and $L_{\mathrm{CE}}^{t}$ is the trigger-only cross-entropy. To explicitly separate the clean and triggered representations of the same image, \textsc{SABLE} uses the margin-based feature-separation loss
\begin{equation}
L_{\mathrm{sep}} =
\Big[
\delta - \|\phi_{\theta}(x_i^c) - \phi_{\theta}(x_i^t)\|_2^2
\Big]_+,
\end{equation}
where $\delta > 0$ is a margin and $[z]_+ = \max\{0,z\}$. To keep the malicious client update close to the received global model $\theta^{(g)}$, it also uses parameter regularization
\begin{equation}
L_{\mathrm{reg}} =
\frac{1}{|S|}
\sum_{j \in S}
\|\theta_j - \theta_j^{(g)}\|_2^2,
\end{equation}
where $S$ is the set of trainable parameters. The overall malicious objective is then
\begin{equation}
L_{\mathrm{mal}} =
L_{\mathrm{CE}}^{\mathrm{tot}} + \lambda_{\mathrm{sep}} L_{\mathrm{sep}} + \lambda_{\mathrm{reg}} L_{\mathrm{reg}}.
\end{equation}
This formulation encourages the model to classify clean inputs correctly, map triggered inputs to the attacker-chosen target class, maintain a feature-level distinction between clean and triggered versions of the same sample, and avoid excessive deviation from the global model. \textsc{SABLE} further incorporates Neurotoxin-style gradient masking so that the backdoor is pushed toward lower-importance parameter subspaces. Overall, \textsc{SABLE} shows that realistic semantic triggers can remain highly effective even under robust aggregation. However, SABLE still treats the trigger itself as a fixed semantic object or transformation \cite{herath2026sable}.

\subsection{Visual Shortcuts, Color Dependence, and the Role of Trigger Appearance}

A related line of computer vision research shows that modern deep networks are highly sensitive to superficial visual cues. Prior work on shortcut learning and texture bias suggests that models often rely on simple visual correlations rather than intended semantics \cite{geirhos2019texture,geirhos2020shortcut}. Other studies show that color is itself an important feature dimension: hue, chroma, and color distortions can materially affect recognition behavior and robustness \cite{flachot2021color,de2021impact,singh2020assessing}. Taken together, these findings suggest that even when trigger geometry, location, and semantics remain fixed, color may change the saliency and learnability of the shortcut signal. This motivates studying trigger color explicitly in federated backdoor settings.

\section{Methodology}
\label{sec:methodology}

\subsection{Controlled Study Design}

In this work, we conduct a controlled study to examine how the color of a semantic trigger influences backdoor behavior in federated learning. Our goal is not to introduce a new attack mechanism, but rather to isolate trigger color as the main experimental factor while keeping the task, trigger semantics, poisoning procedure, and federated training pipeline fixed. The overall experimental pipeline is illustrated in Fig.~\ref{fig:experiment_setup}, while example trigger realizations are shown in Fig.~\ref{fig:trigger_examples}. Building on the same semantics-driven backdoor framework used in our main work \cite{herath2026sable}, we instantiate each trigger in multiple color variants under an otherwise identical setup. This design enables us to study whether different visual realizations of the same semantic trigger lead to different backdoor outcomes under the same federated learning conditions.

\subsection{Federated Learning Task}

We consider a federated hair-color classification setting in which clients collaboratively train a global model through local updates and server-side aggregation. A smaller fraction of clients are malicious, while the larger fraction are benign. Benign clients perform standard local training using only clean data. Malicious clients inject a semantic backdoor by applying a trigger to selected source-class images, relabeling those samples to an attacker-chosen target class, and optimizing their local updates to preserve benign-task utility while inducing targeted misclassification on triggered inputs. This setup follows the general semantics-aware backdoor pipeline of \textsc{SABLE}, enabling a controlled study of trigger-color effects without altering the underlying attack mechanism.

In the defended setting, the malicious objective additionally encourages separation between clean and triggered representations and constrains the poisoned update to remain closer to the evolving global model, thereby improving robustness against server-side defenses. Under this framework, we study whether changing only the color of the semantic trigger leads to different backdoor outcomes across source--target attack directions. Concrete experimental configurations and parameter choices are deferred to Sec.~\ref{sec:experiments}.

\begin{table*}[t]
\centering
\captionsetup{justification=centering}
\caption{\textbf{Mean $\pm$ standard deviation of clean accuracy (Acc) and attack success rate (ASR) for mask and sunglass triggers with black and white color variants in the standard backdoor poisoning setting. Boldface indicates the higher ASR within each black/white trigger-color pair.}}
\label{tab:trigger_performance}
\renewcommand{\arraystretch}{1.2}
\begin{tabular}{|c|c|c|c|c|c|}
\hline
Trigger & Source Hair Color & Target Hair Color & Trigger Color & Acc & ASR \\
\hline
\multirow{4}{*}{Mask} & \multirow{2}{*}{Black} & \multirow{2}{*}{Blond} & Black & $88.62 \pm 1.01$ & $84.70 \pm 11.10$ \\
\cdashline{4-6} 
& & & White & $88.03 \pm 0.51$ & $\mathbf{91.70 \pm 2.05}$ \\
\cline{2-6}
& \multirow{2}{*}{Blond} & \multirow{2}{*}{Black} & Black & $87.84 \pm 0.29$ & $\mathbf{99.90 \pm 0.30}$ \\
\cdashline{4-6} 
& & & White & $86.54 \pm 0.41$ & $91.30 \pm 4.20$ \\
\hline
\multirow{4}{*}{Sunglass} & \multirow{2}{*}{Black} & \multirow{2}{*}{Blond} & Black & $88.31 \pm 0.75$ & $87.90 \pm 4.35$ \\
\cdashline{4-6} 
& & & White & $88.78 \pm 0.54$ & $\mathbf{94.10 \pm 2.33}$ \\
\cline{2-6}
& \multirow{2}{*}{Blond} & \multirow{2}{*}{Black} & Black & $88.14 \pm 0.52$ & $\mathbf{96.60 \pm 1.43}$ \\
\cdashline{4-6} 
& & & White & $88.90 \pm 0.90$ & $93.00 \pm 1.10$ \\
\hline
\end{tabular}
\end{table*}

\begin{table*}[t]
\centering
\captionsetup{justification=centering,singlelinecheck=false}
\caption{\textbf{Mean $\pm$ standard deviation of clean accuracy (Acc) and attack success rate (ASR) for \textsc{SABLE}+defense with mask and sunglass triggers using black and white color variants. Boldface indicates the higher ASR within each black/white trigger-color pair.}}
\label{tab:sable_defence}
\renewcommand{\arraystretch}{1.2}
\begin{tabular}{|c|c|c|c|c|c|}
\hline
Trigger & Source Hair Color & Target Hair Color & Trigger Color & Acc & ASR \\
\hline
\multirow{4}{*}{Mask} & \multirow{2}{*}{Black} & \multirow{2}{*}{Blond} & Black & $78.67 \pm 12.63$ & $71.80 \pm 35.49$ \\
\cdashline{4-6}
& & & White & $83.87 \pm 3.32$ & $\mathbf{86.10 \pm 11.12}$ \\
\cline{2-6}
& \multirow{2}{*}{Blond} & \multirow{2}{*}{Black} & Black & $87.03 \pm 1.65$ & $\mathbf{99.80 \pm 0.40}$ \\
\cdashline{4-6}
& & & White & $86.15 \pm 1.73$ & $95.50 \pm 0.81$ \\
\hline
\multirow{4}{*}{Sunglass} & \multirow{2}{*}{Black} & \multirow{2}{*}{Blond} & Black & $86.68 \pm 1.77$ & $90.10 \pm 2.55$ \\
\cdashline{4-6}
& & & White & $86.25 \pm 1.12$ & $\mathbf{96.90 \pm 2.21}$ \\
\cline{2-6}
& \multirow{2}{*}{Blond} & \multirow{2}{*}{Black} & Black & $86.20 \pm 0.99$ & $\mathbf{97.70 \pm 1.19}$ \\
\cdashline{4-6}
& & & White & $87.12 \pm 1.72$ & $94.90 \pm 1.22$ \\
\hline
\end{tabular}
\end{table*}

\section{Experimental Setup}
\label{sec:experiments}

\subsection{Dataset and Task}

We conduct our study on CelebA and formulate a four-class hair-color classification task with labels black, blond, brown, and gray. All images are resized to $224 \times 224$ and normalized before training. In this workshop study, we consider two targeted attack directions: black hair misclassified as blond hair and blond hair misclassified as black hair. These two directions allow us to examine whether the effect of trigger color remains consistent across opposite source--target mappings rather than being tied to a single label direction.

\subsection{Federated Configuration}

The federated system contains $K=10$ clients, among which two are malicious and the remaining eight are benign. Each client holds 1,000 local training samples. For each malicious client, we construct a poisoned local dataset consisting of 300 triggered samples, 300 paired clean counterparts of those triggered images, and 400 additional clean samples. Benign clients train only on clean data. This design preserves the same local data budget across clients while giving malicious clients access to the clean-triggered pairs required by the \textsc{SABLE} objective.

For evaluation, we use a clean test set and a separate triggered test set. The data preparation pipeline additionally checks for train-test leakage by verifying that clean test samples do not overlap with training data and that triggered test samples do not overlap with the triggered training pool.

\subsection{Trigger Variants}

We study two semantic trigger objects, \emph{masks} and \emph{sunglasses}, and instantiate each in two colors, black and white. This yields four trigger variants:
\[
\{\text{black mask},\ \text{white mask},\ \text{black sunglasses},\ \text{white sunglasses}\}.
\]

We choose black and white as a simple, high-contrast pair of realistic accessory colors. This lets us vary trigger color while keeping the semantic object fixed, and then empirically test whether trigger color interacts with the black--blond source--target directions considered in our experiments. To generate triggered samples, we use \emph{MGIE}, an instruction-based image editing model, to add the desired accessory with the specified color. When needed, the edit is guided toward the relevant facial region so that the accessory is integrated naturally rather than pasted as a synthetic overlay. Images with obvious artifacts are discarded, and retained edited images are preprocessed in the same way as clean images.

For each source-target direction, all trigger variants are generated from the same clean-image pool, and each triggered image is matched to its original CelebA image by image ID. This keeps the comparison controlled: for a fixed attack direction, the intended differences are limited to the trigger object and its color, while identity, pose, background, and the rest of the scene remain as consistent as possible.

\subsection{Training Protocol}

We evaluate two settings to compare the effect of trigger color in a standard attack pipeline and in a stronger defense-aware setting. In the first, malicious clients optimize only clean and triggered cross-entropy losses. In the second, we use the full \textsc{SABLE} objective together with a server-side defense to test whether the same color effect persists under robust aggregation.


Our current implementation uses a ResNet18 backbone, 200 communication rounds, and one local epoch per round. Optimization uses SGD with learning rate $0.001$, momentum $0.9$, and weight decay $5\times10^{-4}$. For the \textsc{SABLE} runs, we set $\lambda_{\text{sep}}=5$ and $\lambda_{\text{reg}}=10^{-6}$. We additionally use Neurotoxin-style gradient masking with preservation ratio $0.05$, and malicious training uses a batch size of 64. In the defended experiments, the server aggregates updates using MultiKrum.

\subsection{Evaluation Metrics}

We report clean accuracy \textbf{\textit{(Acc)}} on the clean test set and attack success rate \textbf{\textit{(ASR)}} on the triggered test set. Acc measures whether the global model preserves benign-task performance (i.e., when the data sample is clean and not the triggered one), while ASR measures how often an image with the trigger is mapped to the attacker-selected target class. Since the goal of this study is to isolate the effect of trigger color, all other components of the federated training pipeline are held fixed across color variants within the same experiment.

\section{Results}
\label{sec:results}

\subsection{Trigger Color Matters Even Without Defense}
Table~\ref{tab:trigger_performance} shows that changing only the color of a semantic trigger can substantially change ASR, even when the trigger object remains fixed. This pattern is visible for both masks and sunglasses and for both attack directions.

When black hair is misclassified as blond hair, the lighter trigger variant consistently performs better. With masks, changing the trigger from black to white increases ASR from \(84.70\%\) to \(91.70\%\) while keeping Acc nearly unchanged (\(88.62\%\) vs.\ \(88.03\%\)). The same trend appears for sunglasses, where white improves ASR from \(87.90\%\) to \(94.10\%\) and slightly improves Acc as well. Thus, when the target class is blond, the white trigger is consistently more effective than its black counterpart.

When blond hair is misclassified as black hair, the same alignment effect is observed. With masks, the black trigger reaches \(99.90\%\) ASR, outperforming the white trigger at \(91.30\%\). With sunglasses, black again performs better, achieving \(96.60\%\) ASR compared to \(93.00\%\) for white. Importantly, these ASR changes do not come with a catastrophic drop in clean performance: across all four trigger variants in the standard setting, Acc remains relatively stable, in a narrow range of \(86.5\%\) to \(88.9\%\).

Taken together, these results show that changing only the color of a semantic trigger can substantially affect ASR even when the trigger object remains fixed. Moreover, the preferred color is not arbitrary: the trigger color that is visually closer to the target class consistently yields higher ASR.


\subsection{The Color Effect Persists Under \textsc{SABLE} + Defense}

Table~\ref{tab:sable_defence} shows that the same trend remains when we move to the stronger \textsc{SABLE}+Defense setting. In our current implementation, this defended setup uses MultiKrum at the server, so the results indicate that the color effect survives robust aggregation rather than disappearing under filtering.

When black hair is misclassified as blond hair, white remains the stronger color for both trigger objects. With sunglasses, white increases ASR from \(90.10\%\) to \(96.90\%\). With masks, the improvement is even larger: white reaches \(86.10\%\) ASR, compared to only \(71.80\%\) for black. The black-mask configuration is also substantially less stable, as reflected by its large standard deviations in both Acc and ASR.

When blond hair is misclassified as black hair, black remains the stronger color for both trigger objects. With sunglasses, the black trigger reaches \(97.70\%\) ASR, compared to \(94.90\%\) for white. With masks, black reaches \(99.80\%\) ASR, again outperforming white at \(95.50\%\). These results show that the preferred trigger color is not specific to the undefended setting; the same directional preference remains after incorporating the stronger \textsc{SABLE} objective and robust aggregation.

Another notable outcome is that strong ASR is still achievable under defense while preserving competitive clean accuracy. Except for the unstable black-mask setting targeting blond hair, the defended runs maintain Acc values mostly in the mid-to-high 80\% range while still achieving ASR above \(90\%\) in most settings. \emph{This reinforces the main security implication of the study: a defense evaluated on only one trigger color may substantially underestimate the threat posed by another color realization of the same semantic trigger.}

\subsection{Key Takeaways}
Across both tables, a consistent pattern emerges: the stronger trigger color tends to align with the target-direction semantics. White triggers are more effective when the target is blond, whereas black triggers are more effective when the target is black. Because this pattern appears for both masks and sunglasses, and in both the standard and defended settings, it is unlikely to be explained by a single trigger object or a single training condition.

This observation has two implications. First, trigger color should be treated as a first-class experimental variable in semantic FL backdoor studies. Second, robustness claims should not be based on a single visual realization of a trigger. Two semantically equivalent triggers can behave very differently simply because their color differs, and this difference can remain visible even under a defense-aware training pipeline such as SABLE+MultiKrum.
\section{Conclusion}

This paper presented a controlled study of trigger color in semantic backdoor attacks on federated learning. By varying only the color of masks and sunglasses while keeping the trigger semantics, poisoning procedure, and federated training pipeline fixed, we showed that trigger color alone can materially affect attack success. Across both the standard poisoning setting and the stronger SABLE-based defended setting, the most effective trigger color consistently aligned with the color of the attacker-chosen target class: white variants were generally stronger when targeting blond hair, whereas black variants were stronger when targeting black hair.

These findings suggest that trigger color is not merely a cosmetic property of a semantic backdoor. Instead, it can shape how effectively the malicious pattern is learned and how well it persists under aggregation and defense. More broadly, the results indicate that evaluating federated backdoor defenses using only a single trigger realization may lead to incomplete or overly optimistic conclusions, since semantically similar triggers can exhibit meaningfully different attack behaviors when their appearance changes.

Future work should extend this analysis to additional trigger types, richer color spaces, other datasets and model families, and a broader range of robust aggregation and defense mechanisms. Such extensions would help clarify how visual trigger properties interact with federated training dynamics and influence the practical risk of semantic backdoor attacks.

\bibliographystyle{IEEEtran}  
\bibliography{refs}           

@inproceedings{mcmahan2017communication,
  title     = {Communication-Efficient Learning of Deep Networks from Decentralized Data},
  author    = {McMahan, H. Brendan and Moore, Eider and Ramage, Daniel and Hampson, Seth and Aguera y Arcas, Blaise},
  booktitle = {Proceedings of the 20th International Conference on Artificial Intelligence and Statistics},
  series    = {Proceedings of Machine Learning Research},
  volume    = {54},
  pages     = {1273--1282},
  year      = {2017}
}

@article{kairouz2021advances,
  title   = {Advances and Open Problems in Federated Learning},
  author  = {Kairouz, Peter and McMahan, H. Brendan and others},
  journal = {Foundations and Trends in Machine Learning},
  volume  = {14},
  number  = {1--2},
  pages   = {1--210},
  year    = {2021},
  doi     = {10.1561/2200000083}
}

@inproceedings{bhagoji2019analyzing,
  title     = {Analyzing Federated Learning through an Adversarial Lens},
  author    = {Bhagoji, Arjun Nitin and Chakraborty, Supriyo and Mittal, Prateek and Calo, Seraphin},
  booktitle = {Proceedings of the 36th International Conference on Machine Learning},
  series    = {Proceedings of Machine Learning Research},
  volume    = {97},
  pages     = {634--643},
  year      = {2019}
}

@inproceedings{bagdasaryan2020how,
  title     = {How To Backdoor Federated Learning},
  author    = {Bagdasaryan, Eugene and Veit, Andreas and Hua, Yiqing and Estrin, Deborah and Shmatikov, Vitaly},
  booktitle = {Proceedings of the 23rd International Conference on Artificial Intelligence and Statistics},
  series    = {Proceedings of Machine Learning Research},
  volume    = {108},
  pages     = {2938--2948},
  year      = {2020}
}

@article{sun2019can,
  title   = {Can You Really Backdoor Federated Learning?},
  author  = {Sun, Ziteng and Kairouz, Peter and Suresh, Ananda Theertha and McMahan, H. Brendan},
  journal = {arXiv preprint arXiv:1911.07963},
  year    = {2019}
}

@inproceedings{xie2020dba,
  title     = {{DBA}: Distributed Backdoor Attacks against Federated Learning},
  author    = {Xie, Chulin and Huang, Keli and Chen, Pin-Yu and Li, Bo},
  booktitle = {International Conference on Learning Representations},
  year      = {2020}
}

@inproceedings{zhang2022neurotoxin,
  title     = {Neurotoxin: Durable Backdoors in Federated Learning},
  author    = {Zhang, Zhengming and Panda, Ashwinee and Song, Linyue and Yang, Yaoqing and Mahoney, Michael W. and Gonzalez, Joseph E. and Kannan, Ramchandran and Mittal, Prateek},
  booktitle = {Proceedings of the 39th International Conference on Machine Learning},
  series    = {Proceedings of Machine Learning Research},
  volume    = {162},
  pages     = {26429--26446},
  year      = {2022}
}

@inproceedings{zhang2023a3fl,
  title     = {{A3FL}: Adversarially Adaptive Backdoor Attacks to Federated Learning},
  author    = {Zhang, Hangfan and Jia, Jinyuan and Chen, Jinghui and Lin, Lu and Wu, Dinghao},
  booktitle = {Advances in Neural Information Processing Systems},
  volume    = {36},
  year      = {2023}
}

@article{gu2019badnets,
  title   = {{BadNets}: Evaluating Backdooring Attacks on Deep Neural Networks},
  author  = {Gu, Tianyu and Dolan-Gavitt, Brendan and Garg, Siddharth},
  journal = {IEEE Access},
  volume  = {7},
  pages   = {47230--47244},
  year    = {2019}
}

@article{chen2017targeted,
  title   = {Targeted Backdoor Attacks on Deep Learning Systems Using Data Poisoning},
  author  = {Chen, Xinyun and Liu, Chang and Li, Bo and Lu, Kimberly and Song, Dawn},
  journal = {arXiv preprint arXiv:1712.05526},
  year    = {2017}
}

@article{turner2019label,
  title   = {Label-Consistent Backdoor Attacks},
  author  = {Turner, Alexander and Tsipras, Dimitris and Madry, Aleksander},
  journal = {arXiv preprint arXiv:1912.02771},
  year    = {2019}
}

@inproceedings{sharif2016accessorize,
  title     = {Accessorize to a Crime: Real and Stealthy Attacks on State-of-the-Art Face Recognition},
  author    = {Sharif, Mahmood and Bhagavatula, Sruti and Bauer, Lujo and Reiter, Michael K.},
  booktitle = {Proceedings of the 2016 ACM SIGSAC Conference on Computer and Communications Security},
  pages     = {1528--1540},
  year      = {2016},
  doi       = {10.1145/2976749.2978392}
}

@inproceedings{wenger2021physical,
  author    = {Wenger, Emily and Passananti, Josephine and Bhagoji, Arjun Nitin and Yao, Yuanshun and Zheng, Haitao and Zhao, Ben Y.},
  title     = {Backdoor Attacks Against Deep Learning Systems in the Physical World},
  booktitle = {Proceedings of the IEEE/CVF Conference on Computer Vision and Pattern Recognition (CVPR)},
  pages     = {6206--6215},
  year      = {2021}
}

@inproceedings{blanchard2017machine,
  title     = {Machine Learning with Adversaries: Byzantine Tolerant Gradient Descent},
  author    = {Blanchard, Peva and El Mhamdi, El Mahdi and Guerraoui, Rachid and Stainer, Julien},
  booktitle = {Advances in Neural Information Processing Systems},
  volume    = {30},
  year      = {2017}
}

@inproceedings{yin2018byzantine,
  title     = {Byzantine-Robust Distributed Learning: Towards Optimal Statistical Rates},
  author    = {Yin, Dong and Chen, Yudong and Kannan, Ramchandran and Bartlett, Peter},
  booktitle = {Proceedings of the 35th International Conference on Machine Learning},
  series    = {Proceedings of Machine Learning Research},
  volume    = {80},
  pages     = {5650--5659},
  year      = {2018}
}

@inproceedings{cao2021fltrust,
  title     = {{FLTrust}: Byzantine-robust Federated Learning via Trust Bootstrapping},
  author    = {Cao, Xiaoyu and Fang, Minghong and Liu, Jia and Gong, Neil Zhenqiang},
  booktitle = {Network and Distributed System Security Symposium (NDSS)},
  year      = {2021},
  doi       = {10.14722/ndss.2021.24434}
}

@inproceedings{nguyen2022flame,
  title     = {{FLAME}: Taming Backdoors in Federated Learning},
  author    = {Nguyen, Thien Duc and Rieger, Phillip and Chen, Huili and Yalame, Hossein and M{\"o}llering, Helen and Fereidooni, Hossein and Marchal, Samuel and Miettinen, Markus and Mirhoseini, Azalia and Zeitouni, Shaza and Koushanfar, Farinaz and Sadeghi, Ahmad-Reza and Schneider, Thomas},
  booktitle = {31st USENIX Security Symposium (USENIX Security 22)},
  pages     = {1415--1432},
  year      = {2022}
}

@inproceedings{geirhos2019texture,
  title     = {ImageNet-trained {CNN}s are Biased Towards Texture; Increasing Shape Bias Improves Accuracy and Robustness},
  author    = {Geirhos, Robert and Rubisch, Patricia and Michaelis, Claudio and Bethge, Matthias and Wichmann, Felix A. and Brendel, Wieland},
  booktitle = {International Conference on Learning Representations},
  year      = {2019}
}

@article{geirhos2020shortcut,
  title   = {Shortcut Learning in Deep Neural Networks},
  author  = {Geirhos, Robert and Jacobsen, J{\"o}rn-Henrik and Michaelis, Claudio and Zemel, Richard and Brendel, Wieland and Bethge, Matthias and Wichmann, Felix A.},
  journal = {Nature Machine Intelligence},
  volume  = {2},
  number  = {11},
  pages   = {665--673},
  year    = {2020},
  doi     = {10.1038/s42256-020-00257-z}
}

@article{flachot2021color,
  title   = {Color for Object Recognition: Hue and Chroma Sensitivity in the Deep Features of Convolutional Neural Networks},
  author  = {Flachot, Alban and Gegenfurtner, Karl R.},
  journal = {Vision Research},
  volume  = {182},
  pages   = {89--100},
  year    = {2021},
  doi     = {10.1016/j.visres.2020.09.010}
}

@inproceedings{de2021impact,
  title     = {Impact of Colour on Robustness of Deep Neural Networks},
  author    = {De, Kanjar and Pedersen, Marius},
  booktitle = {Proceedings of the IEEE/CVF International Conference on Computer Vision Workshops (ICCVW)},
  pages     = {21--30},
  year      = {2021},
  doi       = {10.1109/ICCVW54120.2021.00009}
}

@article{singh2020assessing,
  title   = {Assessing The Importance Of Colours For {CNNs} In Object Recognition},
  author  = {Singh, Aditya and Bay, Alessandro and Mirabile, Andrea},
  journal = {arXiv preprint arXiv:2012.06917},
  year    = {2020}
}

@misc{herath2026sable,
      title={Beyond Corner Patches: Semantics-Aware Backdoor Attack in Federated Learning}, 
      author={Kavindu Herath and Joshua Zhao and Saurabh Bagchi},
      year={2026},
      eprint={2603.29328},
      archivePrefix={arXiv},
      primaryClass={cs.CR},
      url={https://arxiv.org/abs/2603.29328}, 
}

@inproceedings{nguyen2023iba,
  title     = {{IBA}: Towards Irreversible Backdoor Attacks in Federated Learning},
  author    = {Nguyen, Thuy Dung and Nguyen, Tuan A. and Tran, Anh and Doan, Khoa D. and Wong, Kok-Seng},
  booktitle = {Advances in Neural Information Processing Systems},
  volume    = {36},
  year      = {2023}
}

@inproceedings{wang2022flare,
  title     = {{FLARE}: Defending Federated Learning against Model Poisoning Attacks via Latent Space Representations},
  author    = {Wang, Ning and Xiao, Yang and Chen, Yimin and Hu, Yang and Lou, Wenjing and Hou, Y. Thomas},
  booktitle = {Proceedings of the 2022 ACM on Asia Conference on Computer and Communications Security},
  year      = {2022}
}

@article{rieger2022deepsight,
  title   = {DeepSight: Mitigating Backdoor Attacks in Federated Learning Through Deep Model Inspection},
  author  = {Rieger, Phillip and Nguyen, Thien Duc and Miettinen, Markus and Sadeghi, Ahmad-Reza},
  journal = {arXiv preprint arXiv:2201.00763},
  year    = {2022}
}

@inproceedings{rieger2024crowdguard,
  title     = {CrowdGuard: Federated Backdoor Detection in Federated Learning},
  author    = {Rieger, Phillip and Krau{\ss}, Torsten and Miettinen, Markus and Dmitrienko, Alexandra and Sadeghi, Ahmad-Reza},
  booktitle = {Network and Distributed System Security Symposium (NDSS)},
  year      = {2024}
}

@inproceedings{fereidooni2024freqfed,
  title     = {FreqFed: A Frequency Analysis-Based Approach for Mitigating Poisoning Attacks in Federated Learning},
  author    = {Fereidooni, Hossein and Pegoraro, Alessandro and Rieger, Phillip and Dmitrienko, Alexandra and Sadeghi, Ahmad-Reza},
  booktitle = {Network and Distributed System Security Symposium (NDSS)},
  year      = {2024}
}

@article{yang2023filterfl,
  title   = {FilterFL: Knowledge Filtering-based Data-Free Backdoor Defense for Federated Learning},
  author  = {Yang, Yanxin and Hu, Ming and Xie, Xiaofei and Cao, Yue and Zhang, Pengyu and Huang, Yihao and Chen, Mingsong},
  journal = {arXiv preprint arXiv:2308.11333},
  year    = {2023}
}

@inproceedings{liu2018trojaning,
  title     = {Trojaning Attack on Neural Networks},
  author    = {Liu, Yingqi and Ma, Shiqing and Aafer, Yousra and Lee, Wen-Chuan and Zhai, Juan and Wang, Weihang and Zhang, Xiangyu},
  booktitle = {Network and Distributed System Security Symposium (NDSS)},
  year      = {2018}
}

@inproceedings{nguyen2020inputaware,
  title     = {Input-Aware Dynamic Backdoor Attack},
  author    = {Nguyen, Tuan Anh and Tran, Anh},
  booktitle = {Advances in Neural Information Processing Systems},
  volume    = {33},
  year      = {2020}
}

@article{nguyen2021wanet,
  title   = {{WaNet} -- Imperceptible Warping-based Backdoor Attack},
  author  = {Nguyen, Anh and Tran, Anh},
  journal = {arXiv preprint arXiv:2102.10369},
  year    = {2021}
}

@inproceedings{liu2020reflection,
  title     = {Reflection Backdoor: A Natural Backdoor Attack on Deep Neural Networks},
  author    = {Liu, Yunfei and Ma, Xingjun and Bailey, James and Lu, Feng},
  booktitle = {Computer Vision -- ECCV 2020},
  pages     = {182--199},
  year      = {2020}
}

@inproceedings{saha2020hidden,
  title={Hidden Trigger Backdoor Attacks},
  author={Saha, Aniruddha and Subramanya, Akshayvarun and Pirsiavash, Hamed},
  booktitle={Proceedings of the AAAI Conference on Artificial Intelligence},
  volume={34},
  number={07},
  pages={11957--11965},
  year={2020}
  }

@article{zhao2025federation,
  title={The federation strikes back: A survey of federated learning privacy attacks, defenses, applications, and policy landscape},
  author={Zhao, Joshua and Bagchi, Saurabh and Avestimehr, Salman and Chan, Kevin and Chaterji, Somali and Dimitriadis, Dimitris and Li, Jiacheng and Li, Ninghui and Nourian, Arash and Roth, Holger},
  journal={ACM Computing Surveys},
  volume={57},
  number={9},
  pages={1--37},
  year={2025},
  publisher={ACM New York, NY}
}

@inproceedings{sharma2023flair,
  title={Flair: Defense against model poisoning attack in federated learning},
  author={Sharma, Atul and Chen, Wei and Zhao, Joshua and Qiu, Qiang and Bagchi, Saurabh and Chaterji, Somali},
  booktitle={Proceedings of the 2023 ACM Asia Conference on Computer and Communications Security},
  pages={553--566},
  year={2023}
}

@INPROCEEDINGS{herathDSNW,
  author={Herath, Kavindu and Mahangade, Suraj and Bagchi, Saurabh},
  booktitle={2025 55th Annual IEEE/IFIP International Conference on Dependable Systems and Networks Workshops (DSN-W)}, 
  title={A Lightweight Reputation-Based Mechanism for Incentivizing Cooperation in Decentralized Federated Learning}, 
  year={2025},
  volume={},
  number={},
  pages={298-304},
  doi={10.1109/DSN-W65791.2025.00079}}

\end{document}